\documentclass[11pt,twoside]{article}  
\usepackage{adassconf}
\begin{document}   
\paperID{O4-1b}
\title{Atlasmaker: A Grid-based Implementation of the Hyperatlas}
\author{R.\ D.\ Williams, S.\ G.\ Djorgovski, M.\ T.\ Feldmann, and J.\ C.\ Jacob}
\affil{California Institute of Technology, Pasadena, California}
\contact{Roy Williams}
\email{roy@cacr.caltech.edu}
\paindex{Williams, R.D.}
\aindex{Djorgovski, S.G.}
\aindex{Feldmann, M.T.}
\aindex{Jacob, J.C.}
\authormark{Williams, Djorgovski, Feldmann, \& Jacob}
\begin{abstract}          
The Atlasmaker project is using Grid technology, in combination with
NVO interoperability, to create new knowledge resources in
astronomy. The product is a multi-faceted, multi-dimensional,
scientifically trusted image atlas of the sky, made by federating many
different surveys at different wavelengths, times, resolutions, polarizations, etc. The Atlasmaker software does resampling and mosaicking of image collections, 
and is well-suited to operate with the Hyperatlas standard. 
Requests can be satisfied via on-demand computations or by accessing
a data cache. Computed data is stored in a distributed virtual
file system, such as the Storage Resource Broker (SRB). 
We expect these atlases to be a new and powerful
paradigm for knowledge extraction in astronomy, as well as a
magnificent way to build educational resources. The system is being
incorporated into the data analysis pipeline of the Palomar-Quest
synoptic survey, and is being used to generate all-sky atlases from
the 2MASS, SDSS, and DPOSS surveys for joint object detection.
\end{abstract}
\section{Introduction}
Often the combination of multiple datasets reveals more knowledge than is present in the components. An optical survey may show clusters of galaxies, and an Xray survey may show emitting sources. The combination, however, reveals that the Xrays come from the space between the galaxies of the clusters, and we infer the hot intracluster gas that was not obvious in either waveband. This phenomenon is known as {\it data federation}. Extracting knowledge from federated catalogs is well-understood; the next logical step is large-scale federated imagery. This means that images are resampled to a common pixel plane, and catalogs extracted from the joint pixel space. 

The set of images being federated might represent the sky at different
wavelengths, times, resolutions, polarizations, etc. Images may have been drived from combining and computing with other images. In many cases there is knowledge in the federated imagery that is not present in the federation of catalogs derived from the same original images.
It is our contention that the value of this federation can outshine the loss of data quality in the resampling;
indeed it is our contention that image federation opens a new
data-centric window on the Universe. [Jacob 2001, Williams 2000].

This paper describes the Atlasmaker software [Williams and Feldmann 2003] that builds mosaics from NVO-compliant image sets -- ie. images that are exposed through the VO Simple Image Access Protocol [IVOA 2003a]. Atlasmaker scavenges a grid for enough computational resources to build terabyte-size atlases, which are then available for federation and data-mining.

Previous papers [Williams 2003a, Williams 2003b] discusses much of the scientific motivation for image federation, and the resulting need for a standard to ensure interoperability. This standard (Hyperatlas) is a discretization of the space of possible projection from sphere to plane. The standard describes sets of {\it pages} that form {\it atlases}, where a page is defined as a projection of sphere to plane that has a distortion-free point (also called the {\it pointing center}). The Hyperatlas standard also specifies discrete values of the scale at which data is rendered on each page, being a power of two arcseconds per pixel.

When image data are resampled to a standard hyperatlas, they will 
automatically be co-registered at the pixel level with other images, even though each may have been computed by different software. Atlasmaker is an example of such resampling software, and it computes, stores, and delivers this kind of federated image, working in a service economy (i.e., grid), possibly on-demand. 

The creation of federated imagery with scientific value requires great
care. The core computation is the resampling algorithm, which degrades
the data to some (hopefully small) extent. The degradation can be
measured by degrading of the effective PSF (de-focus), an astrometric
shift at the subpixel level, and/or a change in total source flux. We
could simply resample to a very fine scale, reducing the degradation,
but then the computing and data-handling requirements are increased --
by the inverse square of the scale. 

Atlasmaker offers a choice of two trusted codes for the resampling kernel: Montage [Prince 2003] from NASA IPAC/JPL/Caltech and Swarp [Bertin 2003] from the
French Terapix project. These offer different advantages in terms of
quality and speed.

These two codes produce mosaics of multiple images. This
requires an averaging procedure on the areas of overlap, and a means
to subtract ``background''. The separating of background from signal
is a challenge in all areas of science, and each of the codes used by
Atlasmaker has their own chosen method for doing this -- see the
references for more information.

\section{Computing Architecture}

\begin{figure}
\epsscale{0.9}
\plotone{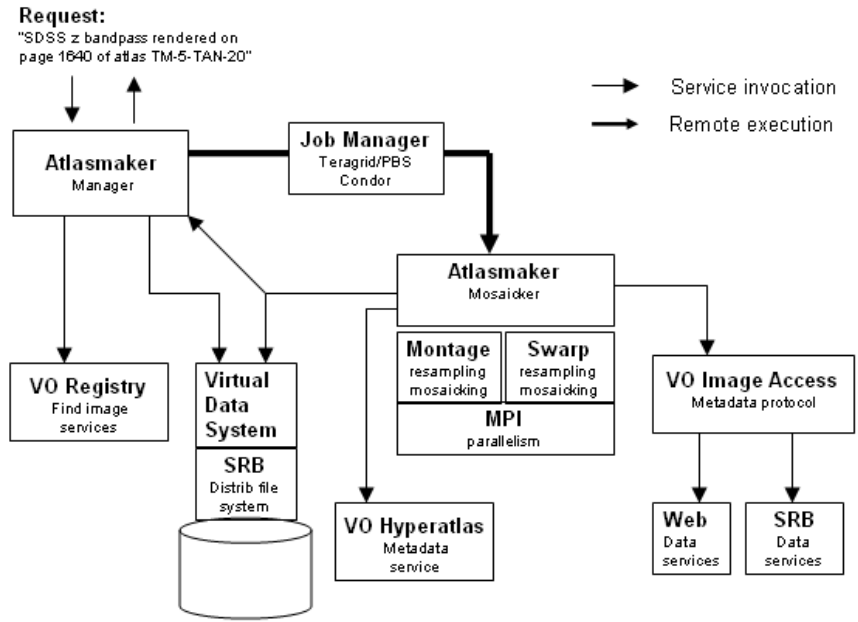}
\caption{Atlasmaker architecture consists of interacting services.} \label{arch}
\end{figure}

Atlasmaker is available on the open-source model [Williams \& Feldmann 2003], and is based on a grid architecture of relocatable programs and web services, as
shown in Figure~\ref{arch}.

When a user requests a given data product, the manager will check for
its existence in the cache that stores already computed products, and
if it is found, it is returned to the user. If it is not found, then
it will be computed. The names of surveys are resolved by the registry
to get required metadata, including current URLs for the NVO services
that can provide external image data.

The Atlasmaker code is written in Python, a highly portable and
powerful scripting language, with a central core of pixel manipulation
written in C (Montage and Swarp). The applications obtain flexibility
through the use of services that may reside anywhere on the net; the
services are replicated for fault tolerance. Data is stored and
retrieved from a distributed file system based on SRB (Storage
Resource Broker), a central pillar of the Teragrid data system.

The job can be wrapped into a self-contained package for execution in
a number of different environments. Currently we have implementations
for the Teragrid queue manager, PBS, and also for the Condor environment.

Atlasmaker has parallel execution modes, allowing resampling of
different images to occur via either message-passing or thread-based
parallelism. Images have background estimated and subtracted, then are
coadded to a mosaic. Completed mosaic images are returned to the user,
but also stored in the data cache in case the same request comes
again.

The Virtual data model (below) encourages use of both on-demand
and scavenger grid utilization. A single mosaic requested by a human
would require on-demand, perhaps anonymous, compute services, whereas
a multi-terabyte sky survey would be resampled over a period of weeks
on whatever resources can be found --- a process known as grid scavenging.

Hyperatlas services are used to convert requested atlas pages to sky
positions. If a user has made a request by sky position, it is
converted first to pages of an atlas, then that page rendered.

For the image data, Atlasmaker uses a standard service type, the IVOA
Simple Image Access Protocol (SIAP) [IVOA, 2003a]. These services can
access data through multiple protocols: current implementations are
ordinary HTTP or the high-performance grid protocol SRB (Storage
Resource Broker). 

Curation metadata from the IVO registries will be used to annotate
computed data, to give it provenance, citations, and other metadata.

\section{Future}

We expect to use Atlasmaker to generate atlas pages both dynamically and in batch mode, the key concept being {\it Virtual Data}. Virtual data as provides a nexus between the worlds of data,
publication, and computing. Data need not be copied and stored, but
another option is available: having the data generated on-demand. A
dataset may be represented by a recipe for its creation, with the
assumption that computing services will be available for its
instantiation when required. That recipe can then be published to a
small or large circle through the Virtual Observatory registry
structure [IVOA 2003b], and hypotheses, excerpts, and conclusions made. 

We expect there to be many sequences of data products stemming from
the original set of mosaicked plates, and we use Virtual Data to
manage them. Virtual Data ensures that derived datasets will respond
in a timely way to changes in the underlying archives, for example
from recalibrations, new data releases, etc. If a mistake is
discovered in an image, we do not want to recompute the entire
multi-terabyte survey again, but only downstream products that are
directly affected.

\acknowledgments
We are grateful to the National Science Foundation for support of this
work through the National Partnerships for Advanced Computational
Infrastructure and through the National Virtual Observatory project. SGD thanks the NASA AISRP program for support. JCJ thanks the NASA for his support.



\begin{references}

\reference Bertin, E., SWarp Resample and Coadd Software, 
\\{\tt \small http://terapix.iap.fr/cplt/oldSite/soft/swarp/index.html}

\reference IVOA, International Virtual Observatory Alliance, 2003a, 
Simple Image Access Protocol, \\
{\tt \small http://us-vo.org/news/simspec.html}

\reference IVOA, International Virtual Observatory Alliance, 2003b,
Registry Framework\\
{\tt \small http://www.ivoa.net/twiki/bin/view/IVOA/IvoaResReg}

\reference Jacob, J.\ C. and Husman, L.\ E.,  Virtual Observatories of the Future Conference, ASP Conference Series, Vol.\ 225, 2001, eds.\ Brunner, R.\ J.,
Djorgovski, S.\ G., and Szalay, A.\ S., pp.\ 192-196.

\reference Prince, T.\ A., Montage Mosaicking Software, 2003, \\{\tt \small
http://montage.ipac.caltech.edu}

\reference Williams, R.\ D., Virtual Sky, 2000, 
{\tt \small http://virtualsky.org}

\reference Williams, R.\ D., The NVO Hyperatlas Standard, 2003a, \\{\tt \small
http://bill.cacr.caltech.edu/usvo-pubs/files/hyperatlas.pdf}

\reference Williams, R.\ D., Djorgovski, S. G., Feldmann, M. T., and Jacob, J. C., Hyperatlas, A New Framework for Image Federation, 2003b, 
astro-ph/0312195.

\reference Williams, R.\ D., Feldmann, M.\ T., Atlasmaker, Software to build Atlases of  Federated Imagery, 2003, \\{\tt \small
http://www.cacr.caltech.edu/projects/nvo/atlasmaker/}
\end{references}
\end{document}